# Dynamics of Successive Minor Hysteresis Loops

Y. W. Windsor, A. Gerber and M. Karpovski

Raymond and Beverly Sackler Faculty of Exact Sciences,

School of Physics and Astronomy

Tel Aviv University, Ramat Aviv 69978 Tel Aviv, Israel

Cumulative growth of successive minor hysteresis loops in Co/Pd multilayers with perpendicular anisotropy was studied in the context of time dependent magnetization reversal dynamics. We show that in disordered ferromagnets, where magnetization reversal involves nucleation, domains' expansion and annihilation, differences between the time dependencies of these processes are responsible for accumulation of nuclei for rapid domain expansion, for the asymmetry of forward and backward magnetization reversals and for the respective cumulative growth of hysteresis loops. Loops stop changing and become macroscopically reproducible when populations of upward and downward nucleation domains balance each other and the respective upward and downward reversal times stabilize.

## Introduction

Reproducibility of cycling processes under identical external conditions attracts particular attention in studies of magnetic hysteresis for at least two reasons: the effect is fundamental to all magnetic storage technologies, and it can serve as a model for a wide range of type-I phase transitions. Two types of hysteresis loops are distinguished. "Major" loops are obtained by sweeping the applied field beyond the saturation field $H_{sat}$, above which the system is said to be fully saturated and its magnetization $M(H)$ is a single-valued function of the applied field $H$; "minor" loops are obtained when the cycle is limited to fields near and below $H_{sat}$. Many studies were devoted to the macroscopic and microscopic repeatability of minor and major hysteresis loops and effects of disorder on cycle-to-cycle memory [1,2]. Macroscopically the major loops repeat themselves in successive cycles, however it was appreciated [3,4] that disorder induces memory in the microscopic magnetic domain configurations from one cycle of the hysteresis loop to the next, even when $H_{sat}$ was crossed. Successive minor loops are usually reasonably reproducible, although changes of size and shape were observed [5-7]. Nevertheless these finite perturbations did not undermine a generally accepted view that macroscopic hysteresis loops are repeatable. It came therefore as a surprise when Berger *et al.* [8] reported major cumulative expansion of successive minor hysteresis loops in Co/Pt and Co/Pd multilayers even up to four times their initial area. The phenomenon was explained in terms of successive accumulation of small nucleation domains, i.e. microscopic areas of reversed magnetization that act as nuclei for the subsequent macroscopic magnetization reversal [8,9]. The reversals at all but the first cycle do not start anew from a fully saturated state but from a preconfigured state in which nuclei for domain expansion are locally conserved. Conservation and accumulation of the nuclei can happen if the system responds asymmetrically in forward and reverse field orientations such that small domains appear to be substantially more stable toward annihilation in an opposite field than toward expansion in a forward field.

In this paper we shall demonstrate that in disordered ferromagnets, where magnetization reversal involves nucleation, rapid domains' expansion and slow annihilation, *differences* between the time dependencies of these processes are responsible for an observed asymmetry of forward and backward magnetization reversals and cumulative growth of minor hysteresis loops.

## Experimental

Co/Pd multilayers exhibiting perpendicular magnetic anisotropy were fabricated by electron beam deposition of 0.2 nm thick Co and 0.9 nm thick Pd multilayers over a room temperature GaAs substrate covered with a 10 nm thick Pd seed layer. The data presented herein were collected at room temperature for a sample with six Co/Pd bilayers, although the results are typical for all tested samples. Hysteresis loops were monitored using the extraordinary Hall effect technique with the field applied perpendicular to the film plane. The Hall voltage in ferromagnetic systems generally follows the relation [10]

$$V_H = \frac{I}{t}\left(R_{OH} B_z + \mu_0 R_{EHE} M_z\right) \tag{1}$$

where $I$ is electrical current, $t$ is the sample thickness, $R_{0H}$ and $R_{EHE}$ are the ordinary and extraordinary Hall effect (EHE) coefficients [11], and $B_z$ and $M_z$ are magnetic field induction and magnetization components normal to the film respectively. $R_{OH}$ is found from measuring $V_H$ in fields beyond the sample's magnetic saturation, and can then be used to subtract the term linear in $B_Z$ from Eq. (1), producing a signal directly proportional to the sample's magnetization $M_Z$:

$$V_{EHE} = \frac{\mu_0 R_{EHE} I}{t} M_z \tag{2}$$

The technique was shown [12, 13] to be highly suitable for measurements of thin magnetic films mainly for two reasons: it is not sensitive to magnetic impurities in the substrate and in the sample's environment, and the measured voltage is significant in thin magnetic films both because of reduction of the film thickness $t$ in Eq. (2) and due to an enhancement of the extraordinary Hall effect coefficient $R_{EHE}$ by surface scattering [14].

Field dependent data were taken at stabilized fields after ramping the field rapidly to the predetermined values. Typical field sweep rate to a target field was about 150 mT/sec. Dwelling time at a given field was never shorter than four seconds.

Hall measurements were done at two geometrical current-voltage configurations $V_{AD,BC}$ and $V_{BC,AD}$, where $V_{AD,BC}$ is voltage measured between contacts B and C with current flowing between A and D, and $V_{BC,AD}$ is voltage measured between contacts A and D when current flows between B and C. Following the reverse magnetic field reciprocity principle [15] $V_{AD,BC}\left(-\vec{M}\right)=V_{BC,AD}\left(\vec{M}\right)$. Therefore, an odd-in-field Hall effect signal, clean from any even-in-field longitudinal resistance component (contributed by an unavoidable Hall voltage contacts' mismatch) can be extracted as

$$V_H = ½\left(V_{\mathrm{BC,AD}} - V_{\mathrm{AD,BC}}\right) \qquad (3)$$

## Results and discussion

### I. Cumulative minor loops

Fig. 1 presents the major and a sequence of minor hysteresis loops observed in the Co/Pd multilayered sample by ramping magnetic field normal to the sample plane. The normalized magnetization is shown as $m_z=\frac{M_z}{M_s}=\frac{V_{EHE}}{V_{EHE,s}}$ where $M_s$ is the saturated magnetization, $V_{EHE}$ is extracted by using Eq. (2) and $V_{EHE,s}$ is the saturated EHE value measured at high fields. The major hysteresis loop was measured by ramping the field between + 1.4 T and -1.4 T such that magnetization is assumed to reach complete saturation in both positive and negative polarizations. The effective field sweep rate was about 150 μT/sec, determined by the number of points taken, dwelling time of 6.7 sec at each measurement point and the ramping time between measurement points. A sequence of minor loops shown here was done by ramping the field between +120 mT and – 120 mT at about 3.2 mT/sec. A higher effective sweep rate was achieved by taking fewer points at larger field intervals. The sequence started from a fully saturated state at +1.4 T. As one can see the resulting minor loops showed a pronounced cumulative growth, advancing deeper and deeper each time the field reached its negative limit at -120 mT. Both the lower and upper bounds of the loop, obtained at the negative and positive limiting fields respectively, evolve with time. However, this evolution is highly asymmetric: the change of restored magnetization at positive field does not exceed 0.5 % of the saturated magnetization $M_s$, while the lower bound developed from 21% of the full reversal magnetization change (equal to $2M_s$) in the first loop to 94% in the 23$^{rd}$ loop. These observations are in a full agreement with results published by Berger et al [8].

## II. **Magnetization reversal at constant magnetic field**

We wish now to demonstrate that time dependent dynamics are an essential ingredient in understanding hysteresis phenomena in general and cumulative minor loops in particular. Fig. 2 presents three major loops measured between fields of + 1.4 T and – 1.4 T at different effective sweep rates: 36 μT/sec, 145 μT/sec and a maximal rate of 150 mT/sec. The maximal rate loop was measured by ramping the magnet from zero field directly to each target field, followed by restoring the magnetic saturation at +1.4 T or -1.4 T. While the width of the magnetization reversal is about 30 mT, roughly the same for all three sweeps, the value of coercive field changes significantly from about 110 mT at a rate of 35 μT/sec to 125 mT at the maximal rate. Reduction of coercive field at lower ramping rates is well understood [16] as a result of thermally-assisted magnetic relaxation taking place as the field is ramped. Sweeps that are much faster than the sample's response time produce wide hysteresis loops. Slower sweeps produce narrow loops. It is therefore important to clarify the connection between minor loops and time dependent reversal phenomena.

Fig. 3(a) presents the time dependence of magnetization reversal from the positively saturated magnetization $M_s$ under different constant negative fields. The measurements were conducted by first saturating the sample in a high positive field and then ramping the field directly to the target field and fixing it constant for the duration of the experiment. For consistency, in the following we define "relaxation" as reversal of magnetization from positive to negative under negative magnetic field. Relaxation is negligible at low fields and grows faster in the vicinity of the coercive field. Full reversal to -$M_s$ was observed within 300 sec at fields exceeding 120 mT. At fields below 90 mT the complete reversal was not achieved even after several hours. Similar to previously studied cases of time dependent relaxation [17,18] one can scale the data by introducing the field dependent parameter $t_{50}^{\downarrow}$ defined as the time required to reverse half of the magnetization at a given field. As shown in Fig. 3b all relaxation data collapse on a single curve when plotted as a function of normalized time $t / t_{50}^{\downarrow}$.

It is common practice to employ the Fatuzzo-Labrune theory [17,19] to interpret magnetization reversal in thin films. The theory uses two microscopic phenomena: nucleation of new reversed domains and their expansion due to domain wall propagation. These are described by the probability of nucleation per unit time *R(H)* [20] and by the effective domain wall velocity *v(H)*, described by [21,22]:

$$R(H) = R_0 \, \mathrm{e}^{-(E_N - 2HM_S V_N)/k_B T}$$
$$v(H) = v_0 \, \mathrm{e}^{-(E_P - 2HM_S V_P)/k_B T} \qquad (4)$$

in which $E_N$ and $E_P$ are characteristic energies for nucleation and propagation, $V_N$ and $V_P$ are characteristic nucleation and propagation volumes (which are assumed to be roughly equal [7]). A typical activation volume $V_a$ can be estimated from an exponential field dependence of $t_{50}$ on the magnetic field [17,21]

$$t_{50} = \tau_0 \exp(-\alpha H) \qquad (5)$$

with $V_a$ given by [20]

$$V_a = \frac{\alpha k_B T}{M_s} \qquad (6)$$

By fitting Eq. (5) to the experimental data, the $\alpha$ coefficient is found to be $16.47\times10^{-2}$ $mT^{-1}$ and $\tau_0$ is $8.84\times10^{9}$ sec. For $M_S = 3\cdot10^5$ A/m [23], $V_a$ is calculated as 2250 $nm^3$, corresponding to an activation radius of 11 nm.

Following the Fatuzzo-Labrune theory, a sample's magnetization reversal can be characterized by a dimensionless parameter *k* defined as

$$k = \frac{v(H)}{R(H)r_n} \tag{7}$$

in which $r_n$ is the radius of a nucleation site. The fraction of *non-reversed* volume is expressed as:

$$\frac{M_Z(t)}{M_S} = 2\exp\left\{-2k^2\left[1-(Rt+k^{-1})+\frac{1}{2}(Rt+k^{-1})^2 - e^{-Rt}(1-k^{-1}) - \frac{1}{2}k^{-2}(1-Rt)\right]\right\} - 1 \ . \tag{8}$$

Two limiting cases can be defined: one in which the nucleation rate dominates the magnetization reversal and $k \ll 1$, and another in which domain wall propagation is dominant and $k \gg 1$. Eq. (8) then reduces to:

$$\frac{M_Z(t)}{M_S} = \begin{cases} 2\exp\left(-k^2(Rt)^3/3\right) - 1 & k \gg 1 \\ 2\exp(-Rt) - 1 & k \ll 1 \end{cases} \tag{9}$$

Fig. 4 presents typical magnetic relaxation data taken at 110 mT. The reversal process cannot be fitted directly to Eq. (8) for the entire time scale. However, the initial stage can be fitted to the $k \gg 1$ limit in Eq. (9) (see dashed line in Fig. 4). The process can therefore be understood as beginning by slow nucleation of small regions with reversed magnetization followed by their rapid expansion by domain wall propagation.

The final stage of the process is important for our discussion. This is when thermally assisted annihilation of the last non-reversed regions takes place [3]. The Fatuzzo-Labrune model does not account for annihilation, however this final part of the process can be well fitted to an expression $M_Z(t) = M_S(A\,\mathrm{e}^{-R_a t} - 1)$, with a prefactor A = 0.63 $M_S$ (see dotted line in Fig. 4). The value of A indicates the stage of the relaxation process at which annihilation becomes significant (at about 70% of reversal). Here we denote $R_a$ as the probability of annihilation per unit time. The calculated values of $R_a$ range between $1.2\times10^{-3}$ Hz at 104 mT and $12.8\times10^{-3}$ Hz at 120 mT. The corresponding effective relaxation times, defined as $\tau_a = 1/R_a$ are 840 sec and 80 sec respectively, about three times longer than $t_{50}$ for their respective fields. Therefore, annihilation is very slow compared to other processes, which will appear to be crucial for understanding the evolution of successive hysteresis loops.

## III. Asymmetric Response to Field Reversal

Cumulative growth of minor loops was associated [8] with asymmetric response to reversal of applied field: it appears to be easier to re-magnetize the system back to the original state (by inverting the magnetic field) than to completely reverse the magnetization by keeping the applied field unchanged. Similar results have been previously reported by Ferré *et al.* [7] for ultrathin cobalt films, and were attributed to the structural defects that give rise to a spatial distribution of local coercive fields that control the domain structure. The "Swiss cheese state" [7] is formed during the domain walls' propagation in which small non-reversed islands located at magnetically harder spots are surrounded by large areas of reversed magnetization. These non-reversed entities act as ready nucleation centers for the remagnetization process, when the magnetic field is subsequently reversed. Therefore a strong difference in dynamic behavior between relaxation in a forward field and remagnetization in a reverse field lies in the presence of non-reversed sites. Thus, when the field is reversed before the system has reached magnetic saturation, the ensuing remagnetization rate is expected to depend on the number of non-reversed domains remaining at the moment of field switching and the magnitude of the positive field applied.

Fig. 5(a) presents a number of remagnetization curves measured at different positive fields applied after 3 minute relaxations at $H_{-}$ = -116 mT. Fig. 5(b) presents the half-reversal times $t_{50}^{\uparrow}$ of these remagnetization processes as a function of applied field. Also shown are half-reversal times found during relaxation from the saturated state, discussed earlier. In both cases the effective reversal time follows the exponential field dependence of Eq. (5). For remagnetization $\alpha = 18\times10^{-2}$ mT$^{-1}$, which is very close to the $16.47\times10^{-2}$ mT$^{-1}$ found previously for relaxation from a saturated state. The calculated activation volume for remagnetization ($V_a$ = 2590 nm$^3$) is very close to that found for relaxation.

Remagnetization is thus interpreted as the same process as the initial relaxation, with the same effective nucleation/propagation volume, but it is much quicker due to the presence of non-reversed enclaves that serve as ready nucleation centers for expansion of upward magnetization. The speed of remagnetization can be reduced to the speed of the initial relaxation only if remagnetization is conducted at a field $H_+$ significantly weaker than the initial relaxation field $H_-$.

The number of remaining non-reversed enclaves is expected to decrease with longer dwelling times. To probe this, we studied the effect of the dwell time in a negative field on the rate of remagnetization. Each experiment began at positive saturation; then a constant negative field $H_-$ was applied for different dwell times allowing magnetization reversal to develop, after which the field was reversed to a positive $H_+$ ($H_+ = |H_-|$). Fig. 6(a) presents the remagnetization curves in positive magnetic field of 116 mT after spending different dwell periods at -116 mT. Fig. 6(b) shows the half reversal times ($t_{50}^{\uparrow}$) of the respective remagnetizations as a function of dwell time in negative field. Remagnetization indeed takes longer when dwell times are longer, so there are less non-reversed sites due to annihilation. Nonetheless, $t_{50}^{\uparrow}$ for all dwell times is much shorter than $t_{50}^{\downarrow}$. This holds even after nearly six hours of dwelling under the negative field ($t_{50}^{\uparrow}$ = 17 sec compared to $t_{50}^{\downarrow}$ = 70 sec).

## IV. Symmetric Minor Loop Evolution

We now turn to study successive minor hysteresis loops (see Fig. 1) as a succession of relaxations and remagnetizations. To produce minor loops we start at positive magnetic saturation $M_S$. The field is ramped rapidly to a negative field $H_-$, in which the sample dwells for $T_-$ (much longer than the field ramp time). The field is then swept back through zero to a positive field $H_+$, where the sample dwells for $T_+$, before the field is taken back to $H_-$. These steps are repeated, creating a square-wave magnetic field with a period of $T = (T_- + T_+)$. This procedure allows singling out four parameters we use to characterize a minor loop ($H_+$, $H_-$ , $T_+$, $T_-$). The total cycle period $T$ is much shorter than time required for accomplishing annihilation of non-reversed sites at these fields (see previous section). We characterize minor loops by their magnetization at the upper and lower bounds, $M_+$ and $M_-$ respectively, and by difference between the two: $\Delta M$.

Fig. 7(a) presents a sequence of symmetric minor loops ($H_+ = |H_-| = 110$ mT and $T_+ = T_- = 74$ sec) along with relaxation data taken at a constant field of the same magnitude starting from saturation. Evidently, $M_-$ evolves towards $-M_s$ and is limited by the constant field relaxation. Fig. 7(b) shows two minor loop sequences measured within $H_+ = |H_-| =$ 110 mT but with different cycle periods $T = 160$ sec and 75 sec. The data are presented as a function of cycle number $n = t/T$. By shortening the cycle period, reaching the same $\Delta M$ requires less cycles. Therefore minor loop evolution depends mainly on the time available for reversal and less on the number of cycles $n$.

Fig. 8 presents the typical evolution of relaxations at successive cycles. The first relaxation is characterized by slow nucleation of reversed domains followed by their expansion through rapid wall motion. In the following cycles the process grows faster and deeper until it becomes reproducible. Both relaxations and remagnetizations in each cycle can be fitted as $M(t) \propto exp(-t/\tau)$. Fig. 9 presents the half-lives $\tau_\downarrow$ and $\tau_\uparrow$ (for relaxation and remagnetization respectively) of these fits as function of cycle number. In the first cycles, $\tau_\downarrow$ and $\tau_\uparrow$ start at very different values (45 sec and 2 sec respectively). It is evident that from cycle to cycle $\tau_\downarrow$ decreases and $\tau_\uparrow$ increases, until they equalize at a common value. Changes in $\tau_\downarrow$ and $\tau_\uparrow$ are most likely due to changing numbers of ready nucleation sites at the beginning of each reversal. Notably, stabilization of $\tau_\downarrow$ and $\tau_\uparrow$ coincides with stabilization of the loop size $\Delta M$ (also shown in Fig. 9), after which further loops become reproducible. The evolution of symmetric minor loops may thus be attributed to the evolution of the number of non-reversed enclaves in both orientations of magnetization that serve as ready nucleation sites for rapid expansion. The higher the number of nucleation sites at the onset of a reversal, the faster it develops and the respective change in magnetization is larger. The loops stop changing and become reproducible when $\tau_\downarrow$ and $\tau_\uparrow$ equalize.

## V. Asymmetric minor loops

It is of further interest to follow the development of minor hysteresis loops under asymmetric conditions: fields $H_+ \neq H_-$ or dwell times $T_+ \neq T_-$. Fig. 10(a) exemplifies a typical asymmetric process in which $|H_-| > H_+$ ($H_-$ = -120 mT and $H_+$ = 110mT). In this case both the lower and the upper bounds ($M_-$ and $M_+$) evolve with cycling significantly. Similar to the symmetric cases previously discussed, $\tau_\downarrow$ and $\tau_\uparrow$ evolve from cycle to cycle. Fig. 10(b) presents $\tau_\downarrow$ and $\tau_\uparrow$ as a function of cycle number $n$, for $H_-$ = -109 mT and $H_+$ =104 mT with $T_+ = T_-$ = 80 sec. Unlike the symmetric-in-field case, $\tau_\downarrow$ and $\tau_\uparrow$ stabilize after a number of cycles but do not *equalize*. The stabilized $\tau_\downarrow$ (under higher field) is shorter than $\tau_\uparrow$. Similar to the symmetric case, $\Delta M$ stabilizes once $\tau_\downarrow$ and $\tau_\uparrow$ cease changing, after which minor loops become macroscopically reproducible.

In the stabilized and reproducible minor loops the magnetization reached at negative and positive fields ($M_-^\infty$, $M_+^\infty$) and the change of magnetization $\Delta M^\infty$ depend on the ratio of the positive and negative applied fields and their corresponding dwell times. Fig. 11(a) illustrates three typical shapes with which minor loops evolve with cycling for $|H_-| < H_+$, $|H_-| = H_+$ and $|H_-| > H_+$ ($T_+ = T_-$). These cases can also be identified in Fig 11(b), which presents $M_-^\infty$, $M_+^\infty$ and $\Delta M^\infty$ as a function of $H_+/|H_-|$. When $H_+ > |H_-|$, $M_+$ remains close to $M_s$ and only $M_-$ changes with cycling. If $H_+$ is strong enough to suppress all negatively magnetized enclaves at each remagnetization event, $M_-$ does not evolve at all, and $\Delta M$ is confined to the change occurring during a single relaxation process under field $H_-$ during time $T_-$. When $H_+ = |H_-|$, $M_+$ remains close to $M_s$ and $M_-$ advances towards $-M_s$. This is the case in which the stabilized hysteresis loop is the largest. Lastly, when $H_+<|H_-|$ both bounds evolve towards $-M_S$ . If $H_+ \ll |H_-|$, the minor loops stabilize such that both $M_-^\infty$, $M_+^\infty$ are close to $-M_S$ and $\Delta M$ is negligible.

In a similar way, one can model hysteresis loops with unequal ramp rates at positive and negative fields by assigning different dwell times $T_+$ and $T_-$. Fig. 12 presents $M_-^\infty$ and $M_+^\infty$ of the stabilized loops as a function of $T_+/T_-$ found for $H_+ = |H_-|$ = 104 mT and $T_-$ =

4 sec. For $T_+ < T_-$ the time allowed for remagnetization is not sufficient to annihilate most of the downward magnetized sites and the sequence stabilizes with $M_+^\infty < M_s$. In the opposite limit of $T_+ \gg T_-$ all the downward magnetized sites are annihilated and each new cycle starts from a fully saturated state. The size of the stabilized hysteresis is then defined by relaxation during $T_-$ under field $H_-$.

## Summary

To summarize, we studied the time and field dependent dynamics of magnetization reversal in Co/Pd multilayers with perpendicular anisotropy. Magnetization reversal proceeds through three stages: nucleation of small reversed sites, their rapid expansion and slow annihilation of remnant non-reversed enclaves. An effective reversal rate is strongly enhanced when ready nucleation centers are already present at the start of the reversal process, which happens when magnetic field is reversed before annihilation is completed. The unequal population of remnant upward and downward magnetized enclaves is the origin of asymmetric reversal rates in positive and negative fields. In particular, this is seen when relaxation initiated at a fully saturated state is followed by a much quicker remagnetization.

Successive minor hysteresis loops were studied as a repetition of fixed field magnetization reversal processes at both field polarities. This simplified case allows understanding the phenomenon of cumulative growth of minor hysteresis loops within the context of magnetization reversal under constant fields - as succession of relaxations and remagnetizations. Successive hysteresis loops were found to evolve according to changes in effective upward $\tau_\uparrow$ and downward $\tau_\downarrow$ reversal times which depend on population of nucleation sites ready at the onset of each process. Hysteresis loops stabilize and become macroscopically reproducible when $\tau_\uparrow$ and $\tau_\downarrow$ stabilize. In the case of symmetric-in-field minor loops, stabilization occurs at $\tau_\uparrow = \tau_\downarrow$. The process can be understood in the framework of a two-directional Swiss-cheese model, in which nucleation sites are generated as small enclaves of non-reversed magnetization left behind the front of propagating domains in both magnetization polarities, regardless of the initial state.

Finally we wish to comment on the question why cumulative changes in successive hysteresis loops observed by Berger et al [8] and by us are so large as compared with previous studies. A simple answer might be in matching the experimental time window and the effective relaxation / remagnetization times. Significant changes are observable from loop to loop for field cycling rates at which the time spent in the vicinity of the coercive field is comparable with $\tau_{\downarrow}$ and $\tau_{\uparrow}$. For higher ramping rates the relative changes between successive cycles are small and one needs to repeat the process many times to observe a significant effect.

## Acknowledgements

This work was supported by the Israel Science Foundation grant No. 633/06 and by the Air Force Office of Scientific Research, Air Force Material Command, USAF grant No. FA8655-11-1-3081.

## Figure Captions

**Fig. 1:** The major (solid line) and a sequence of minor hysteresis loops (●) observed in the $[Co_{0.2}/Pd_{0.9}]_6$ sample by ramping magnetic field normal to the sample plane. Numbers indicate the order of cycles in the sequence and arrows the direction of field ramp.

**Fig. 2:** Major loops measured at different sweep rates: 35 μT/sec (●), 145 (○) μT/sec and 150 mT/sec (▲). The field axis is cut for clarity.

**Fig. 3:** (a) Time dependent magnetization reversal under different constant fields; (b) the same data with time normalized by the corresponding half-reversal time $t_{50}^{\downarrow}$. All reversals start at full saturation.

**Fig. 4:** Magnetization reversal under a 110 mT field as a function of $t/t_{50}^{\downarrow}$. Open circles indicate the measured data. The dotted line is an exponential fit to the final stage of the process and the dashed line is a fit to Eq. (9) in the $k \gg 1$ limit.

**Fig. 5:** (a) Time dependence of magnetization relaxation under an applied field of -115 mT, followed by field reversal and remagnetization under different positive applied fields. The field was switched from negative to positive at *t=0* on the time axis, 3 min after start of relaxation. (b) Half-reversal times of relaxation $t_{50}^{\downarrow}$ and remagnetization $t_{50}^{\uparrow}$ as a function of applied field.

**Fig. 6:** (a) Time dependence of remagnetization in 119 mT after different dwell times in -116 mT; (b) half reversal times $t_{50}^{\uparrow}$ of remagnetization as a function of dwell time.

**Fig. 7:** (a) Succession of symmetric minor loops measured between $H_+ = |H_-|$ = 109 mT (○) and a constant field relaxation process in -109 mT (●) as a function of time. Cycle period $T$ = 234 sec. (b) Two sequences of minor loops measured with periods 235 (●) and 390 (○) between $H_+ = |H_-|$ = 109 mT as a function of cycle number *t/T*.

**Fig. 8:** Relaxations from Fig. 7a as function of cycle time, starting from the beginning of each cycle. Numbers indicate the order of cycles in the sequence.

**Fig. 9**: Half-lives of relaxations (○) and remagnetizations (●) as function of cycle number from the evolution process in Fig. 7a. ΔM is also shown (▲) as function of cycle number.

**Fig. 10**: (a) Example of asymmetric-in-field minor loop evolution ($H_+$ = 110 mT, $H_-$ = -120 mT) with symmetric dwell times of 4 sec; (b) half-life evolution of a similar process

with asymmetric fields ($H_-$ = -109 mT, $H_+$ = 104 mT) and symmetric dwell times of 78 sec. $\Delta M$ is also shown as function of cycle number.

**Fig. 11:** (a) Upper (○) and lower (●) limits of stable reproducible minor loops ($M_-^\infty$, $M_+^\infty$) as function of the ratio $H_+/H_-$ used. Also plotted is their difference $\Delta M^\infty$ (▲). $H_-$ = -120 mT in measurements. (b) Three typical shapes for minor loop sequences, depending of the relation between $H_-$ and $H_+$.

**Fig. 12:** Upper (○) and lower (●) limits of stable reproducible minor loops ($M_-^\infty$, $M_+^\infty$) as function of $T_+/T_-$. Also plotted is their difference $\Delta M^\infty$ (▲). Symmetric fields of 104 mT are applied, and $T_-$ is 4 sec in all measurements.

# Figures

Figure 1

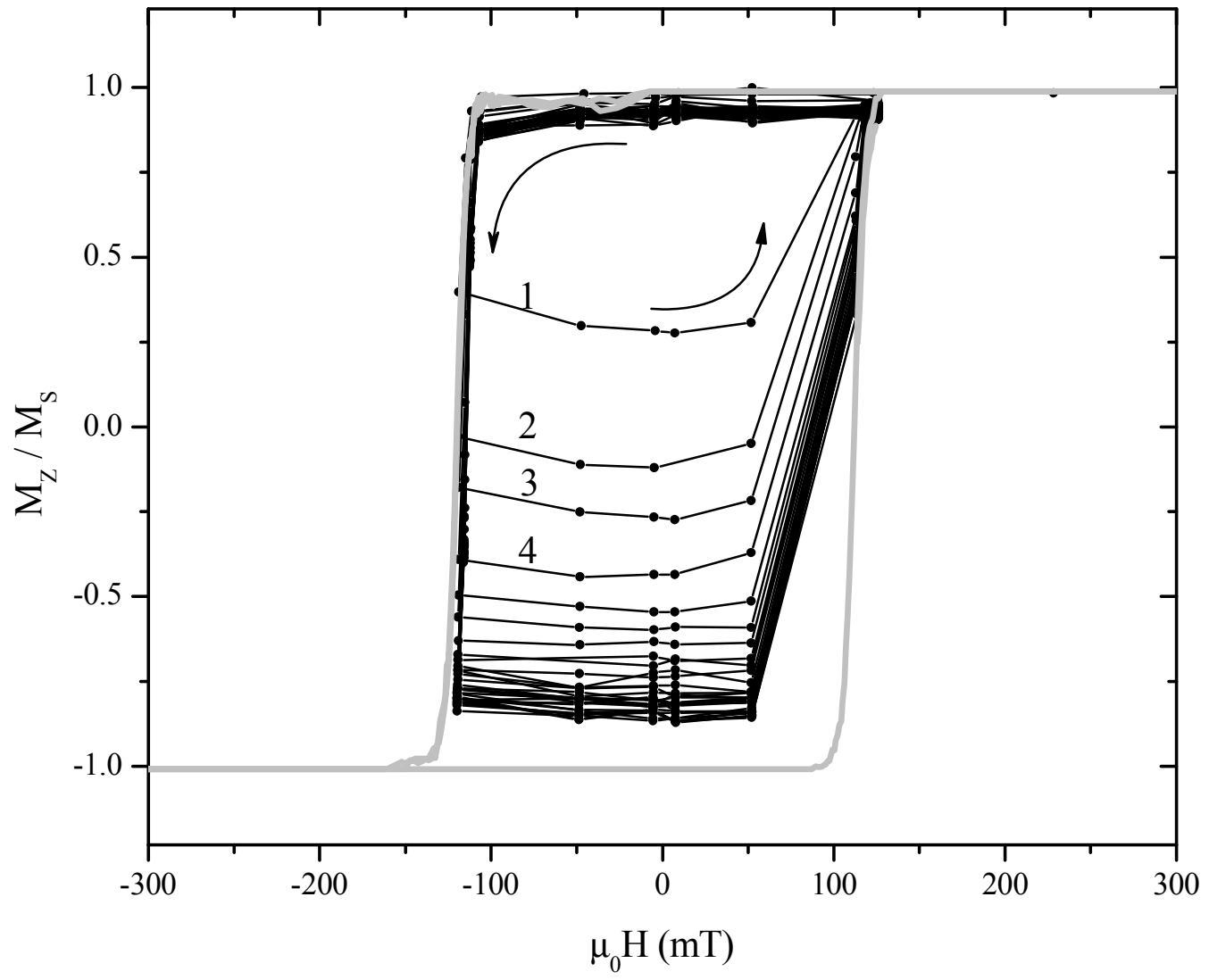

Figure 2

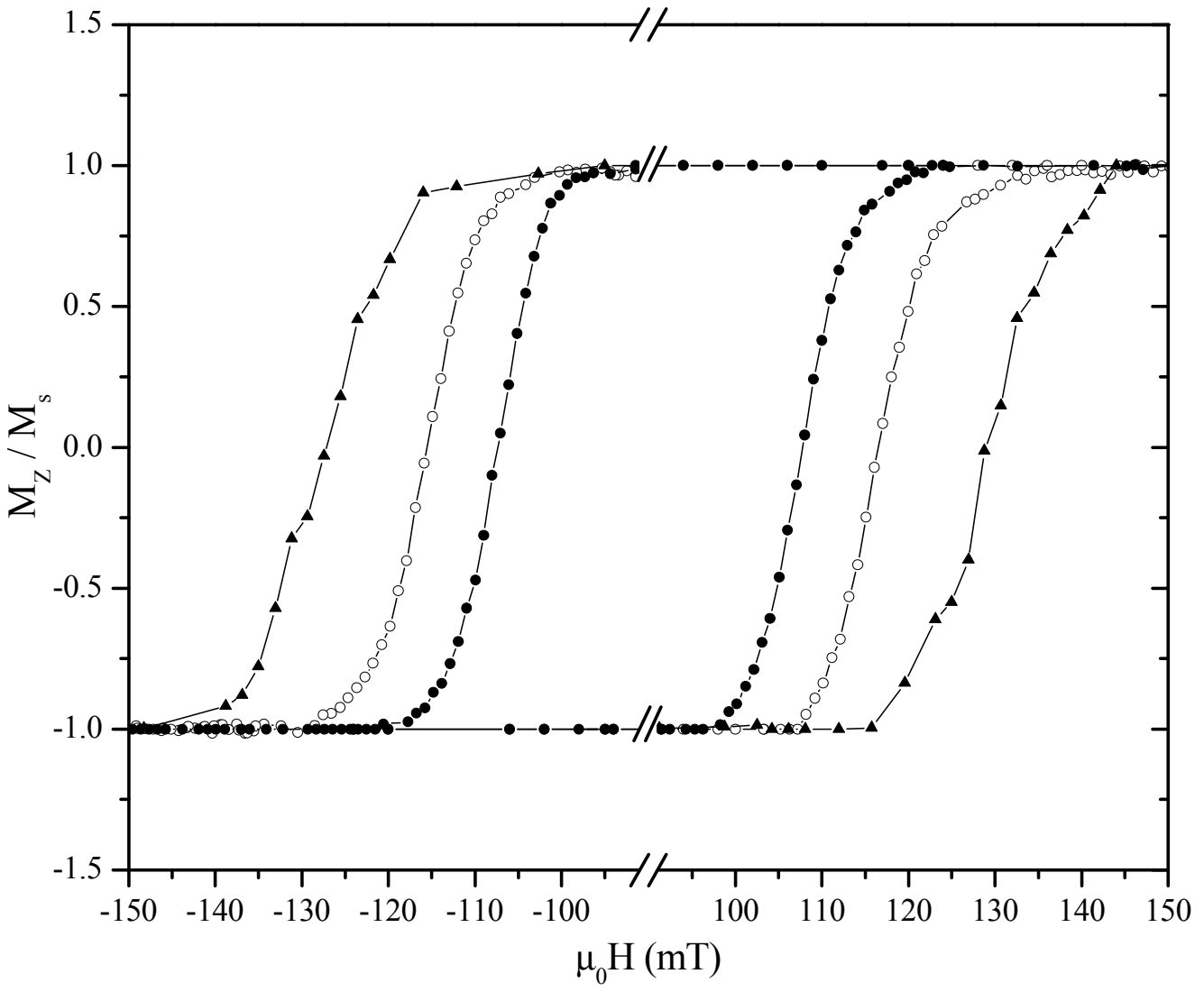

Figure 3

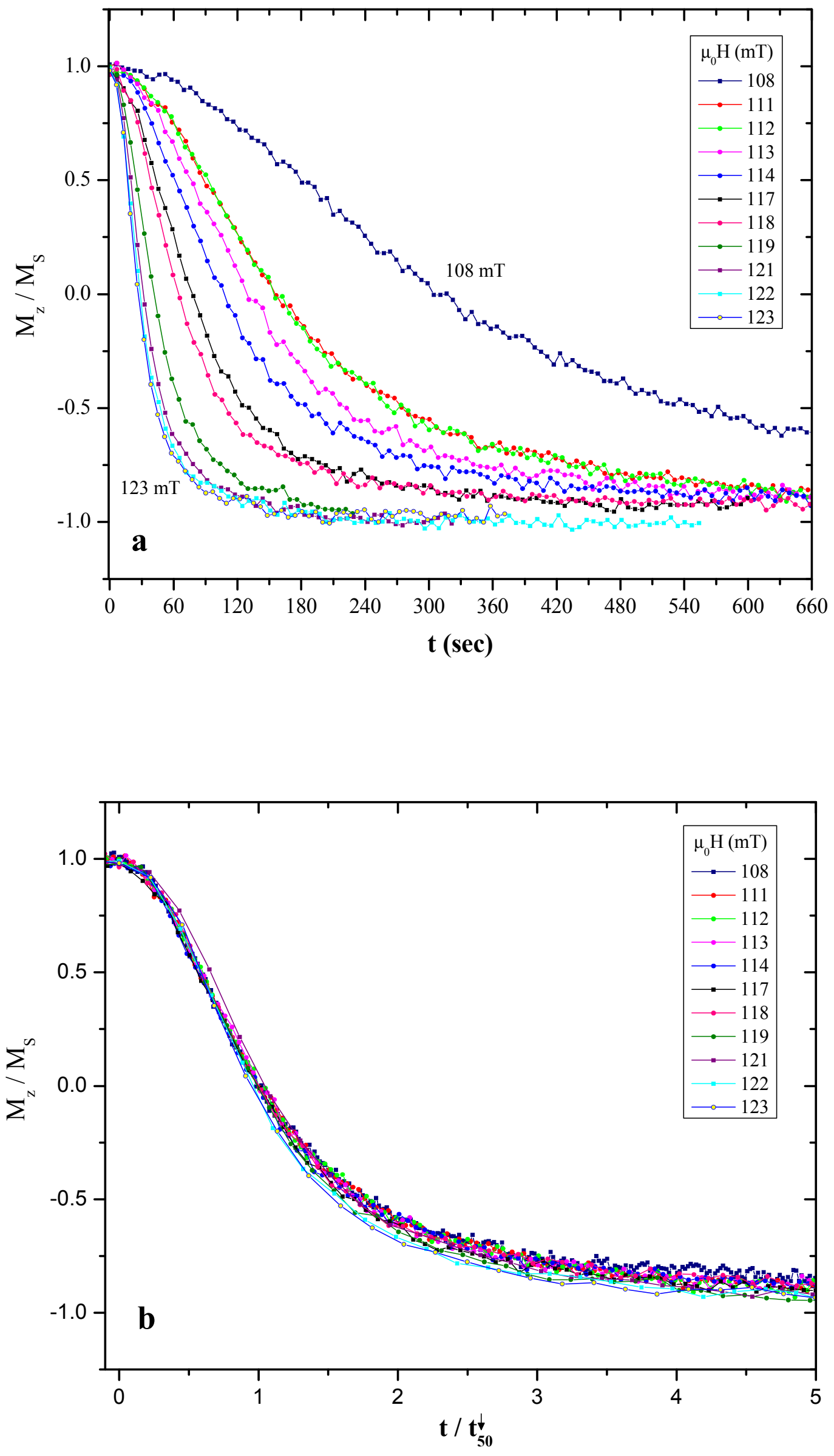

Figure 4

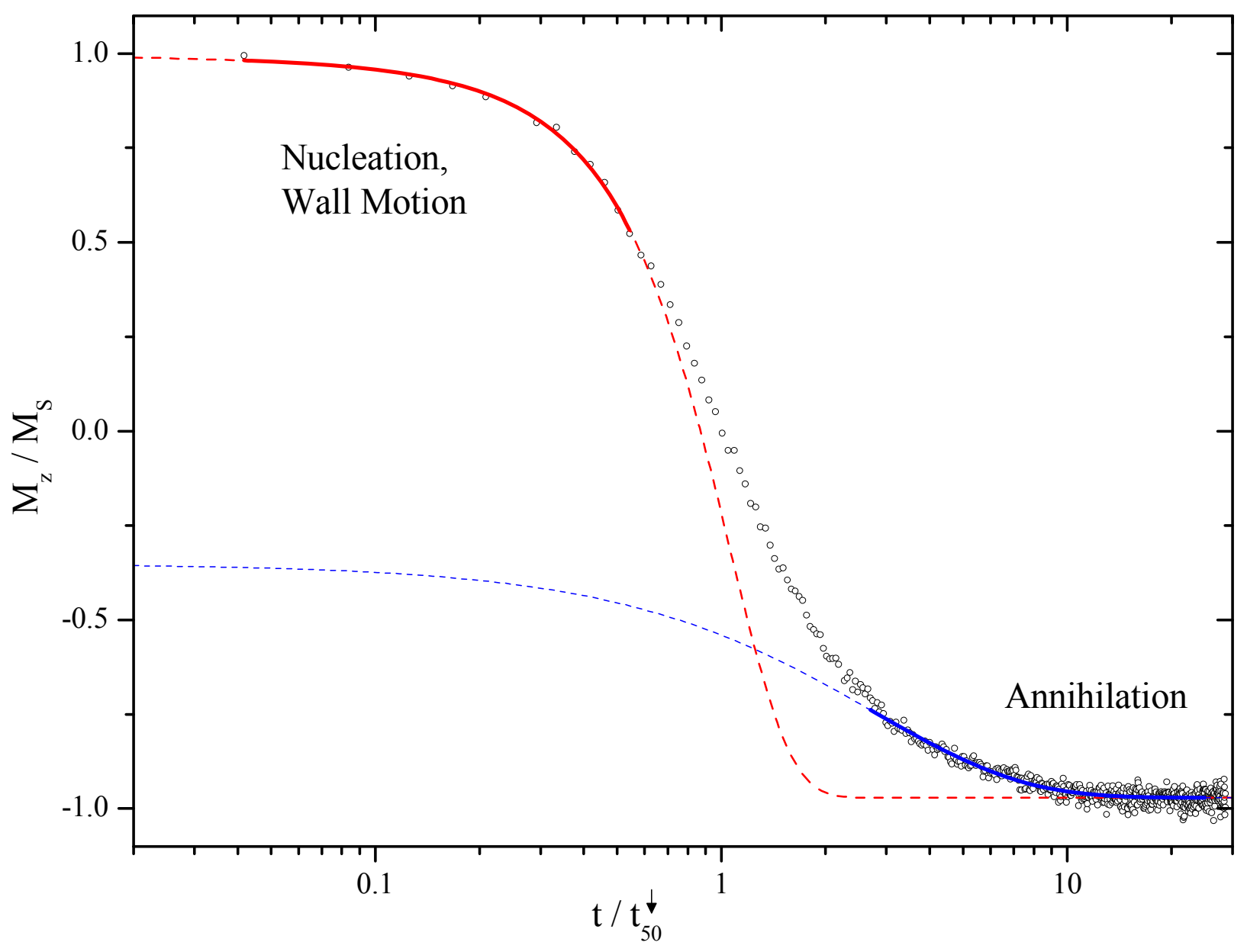

Figure 5

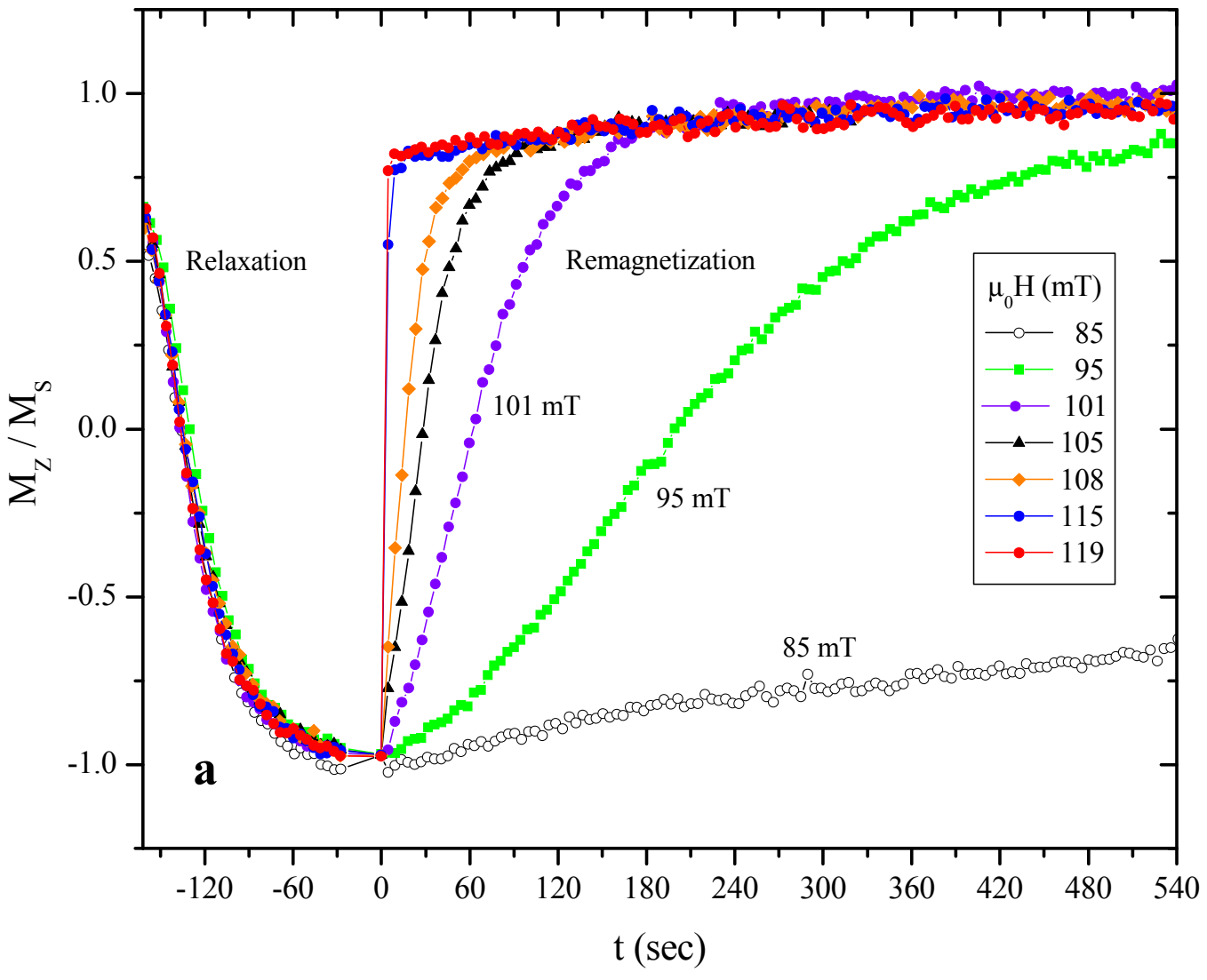

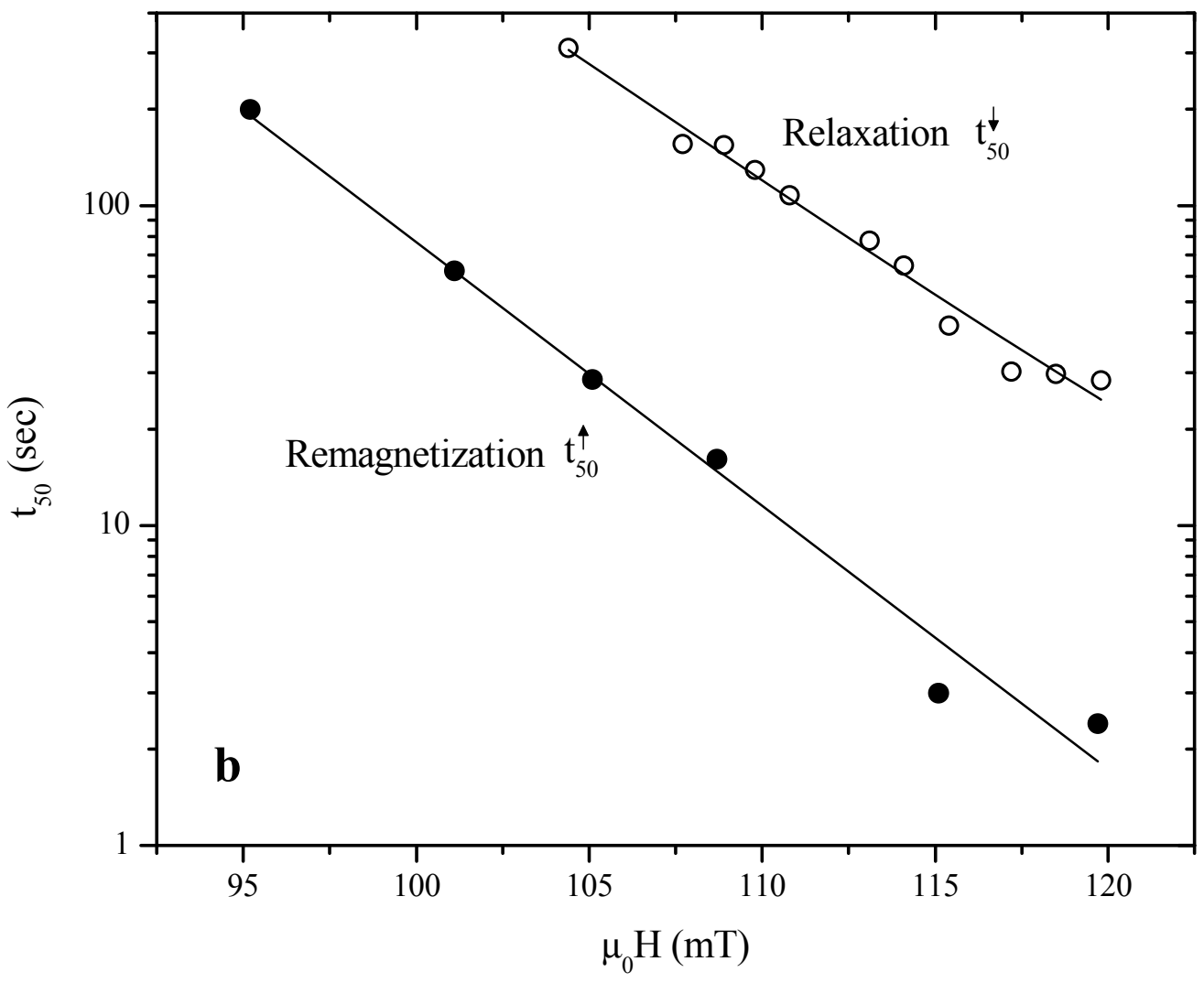

Figure 6

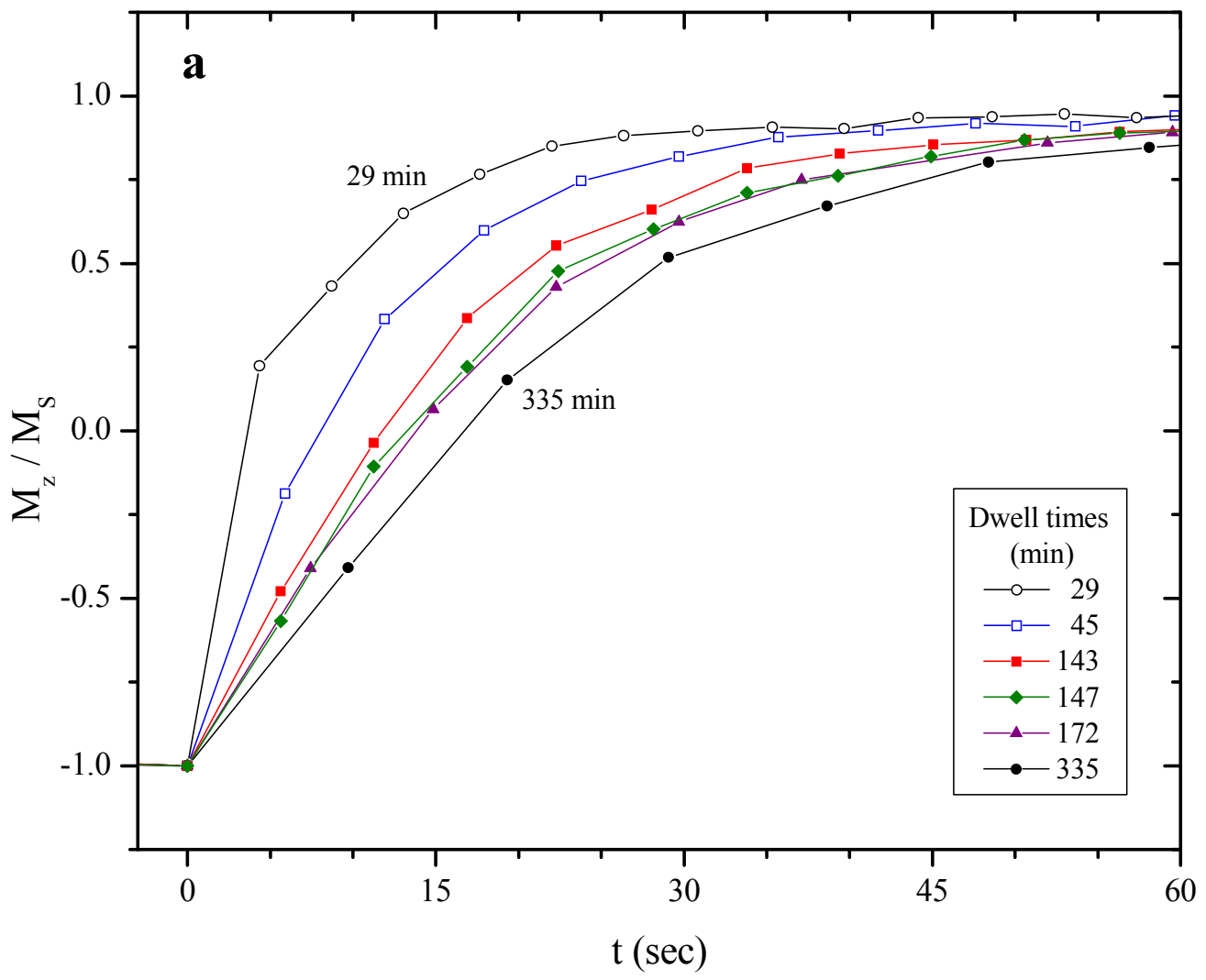

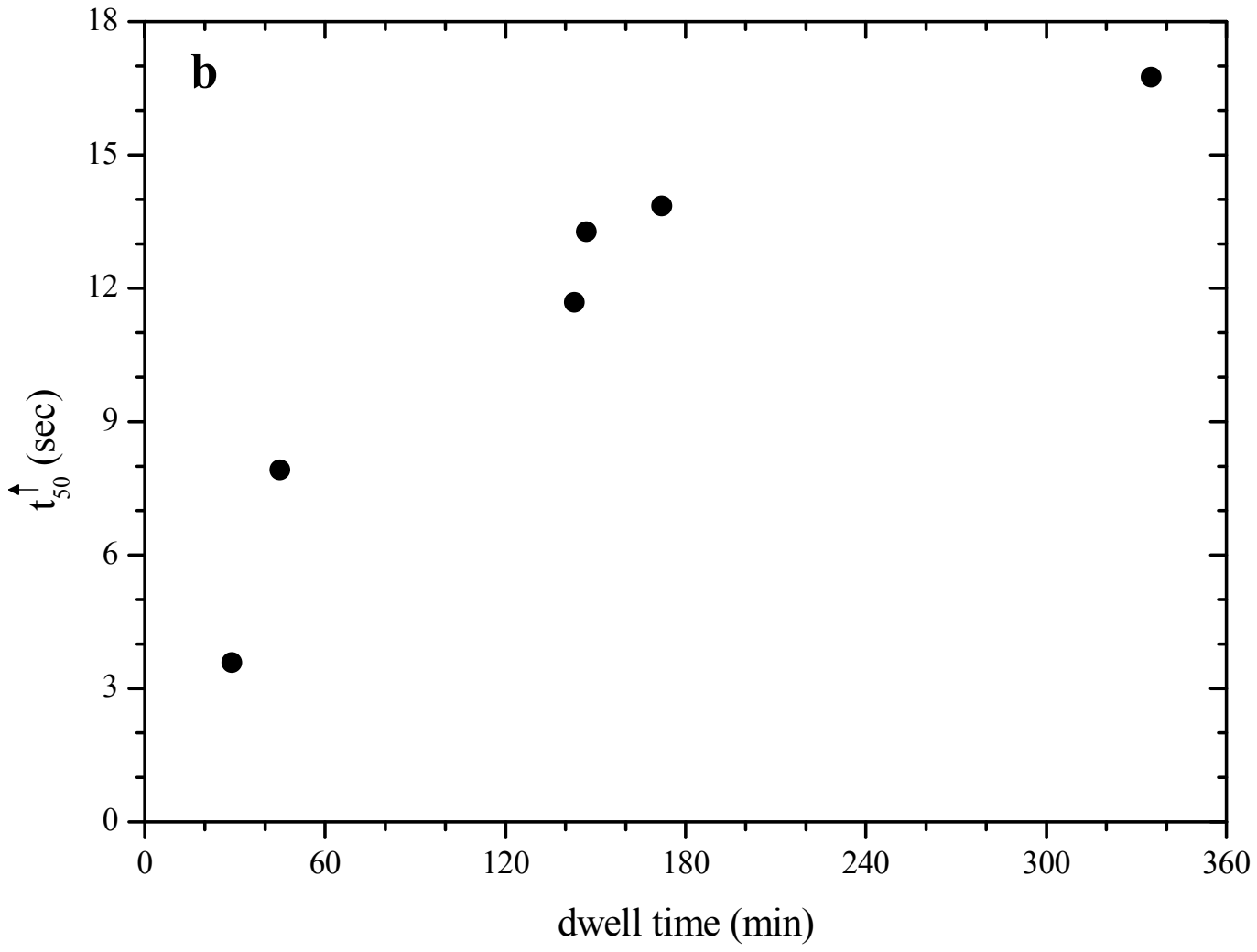
b
$t^{\uparrow}_{50}$ (sec)
dwell time (min)

Figure 7

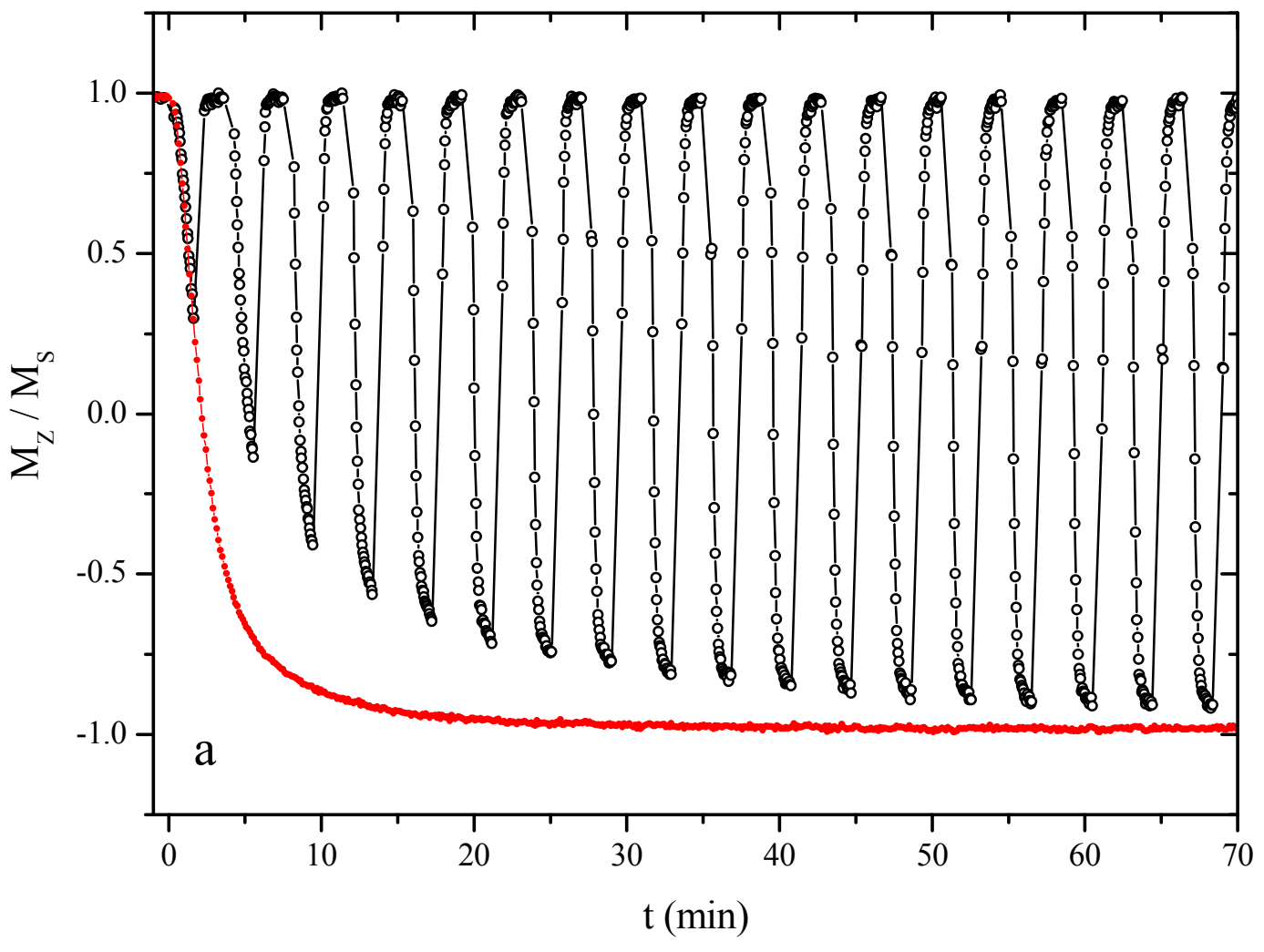

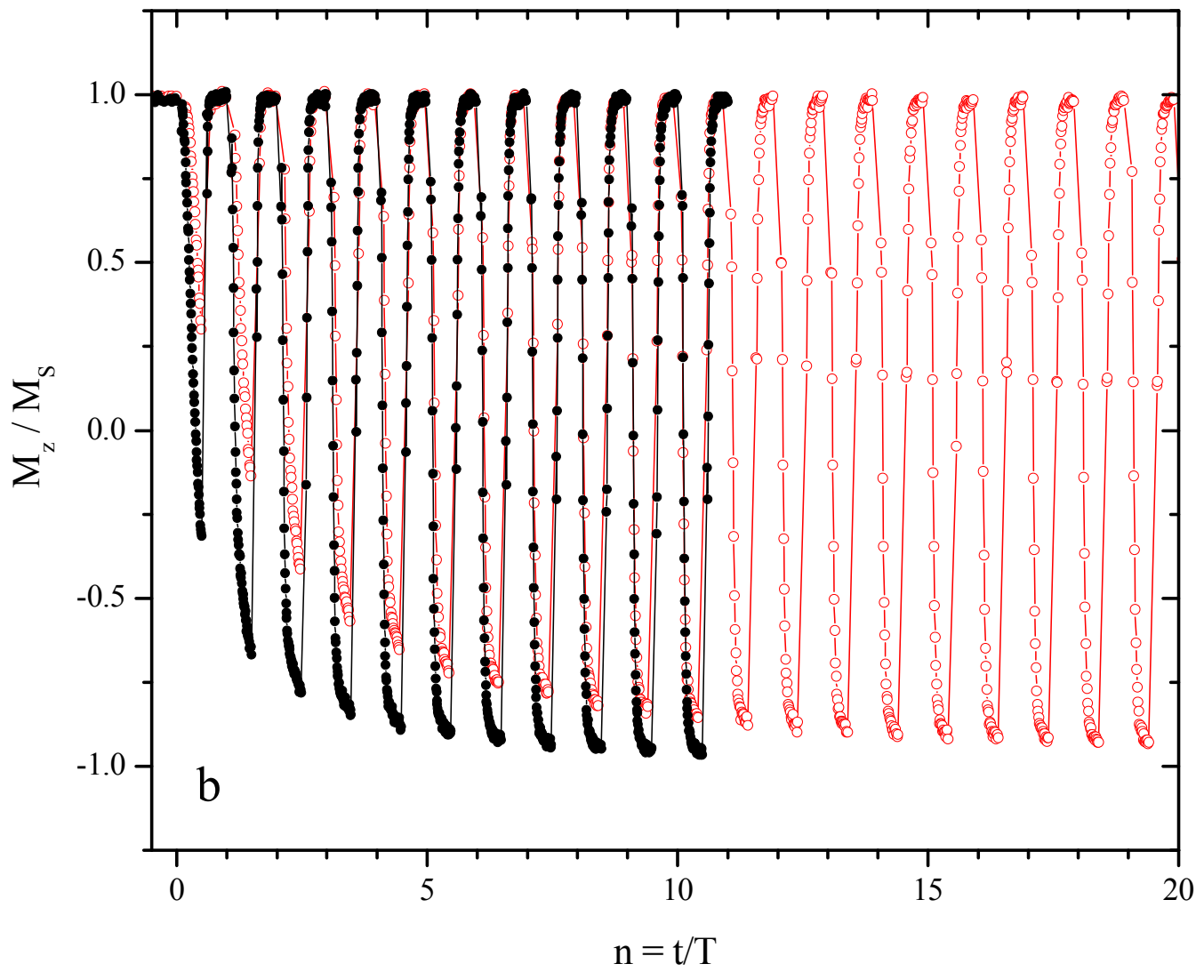

Figure 8

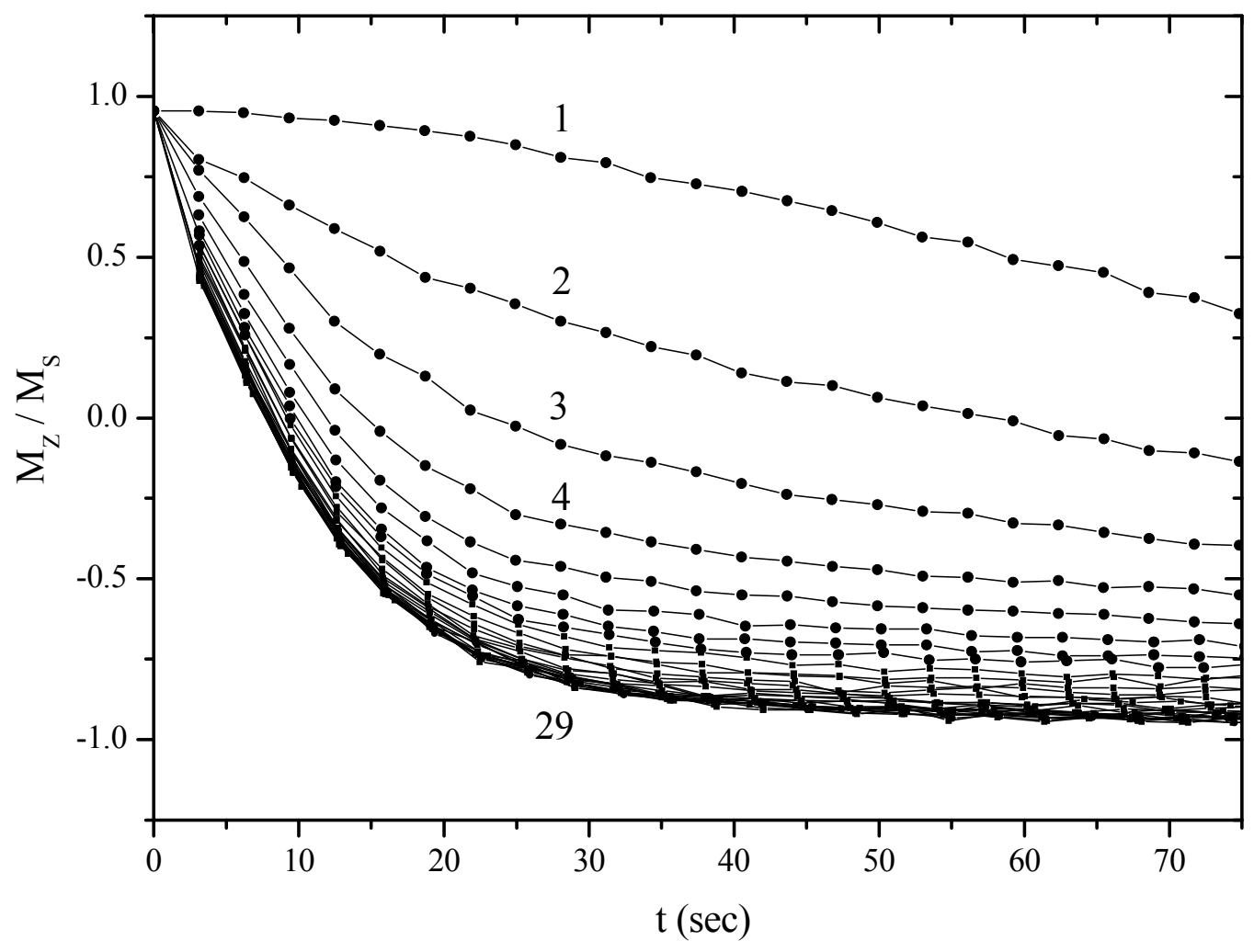

Figure 9

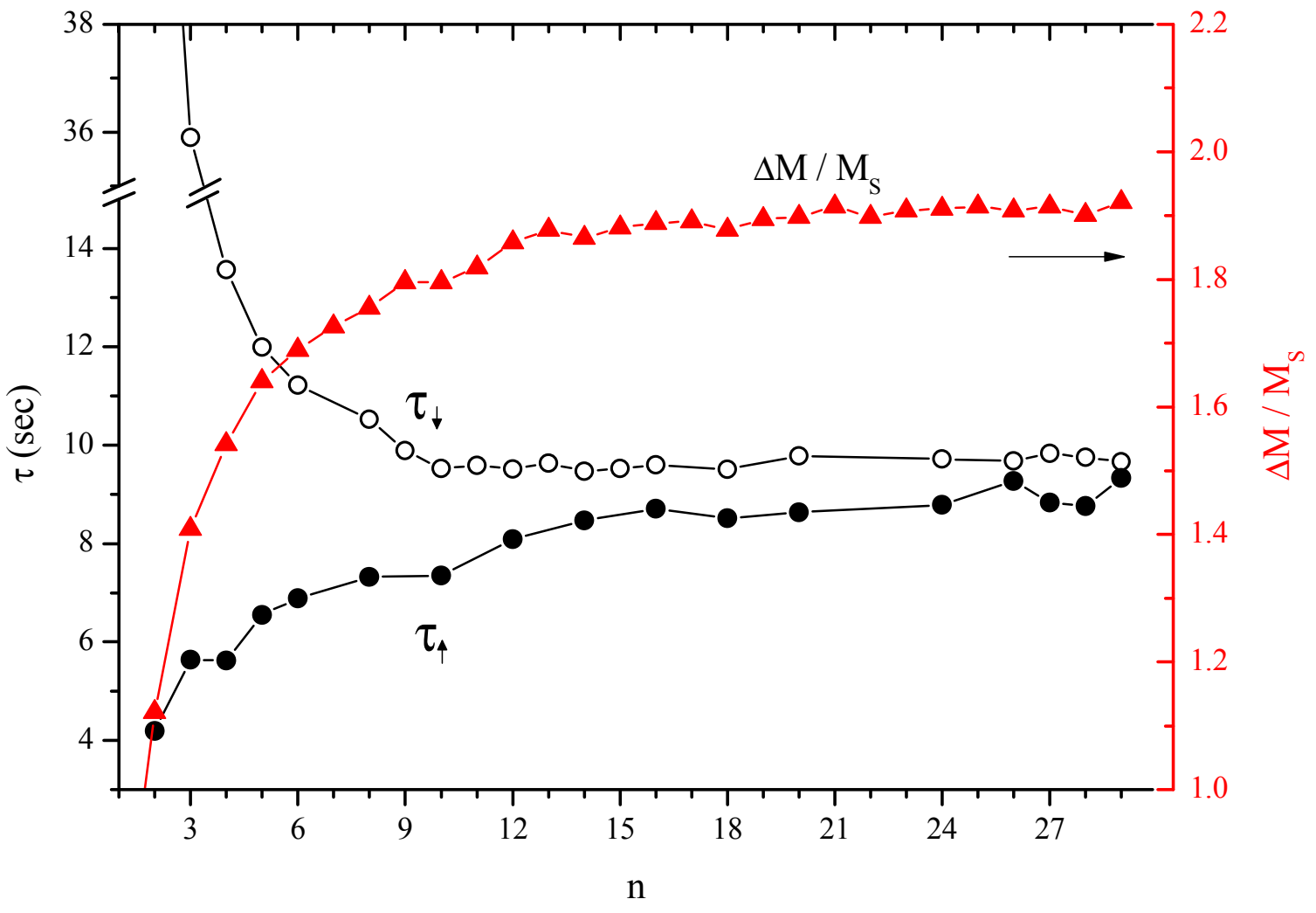

Figure 10

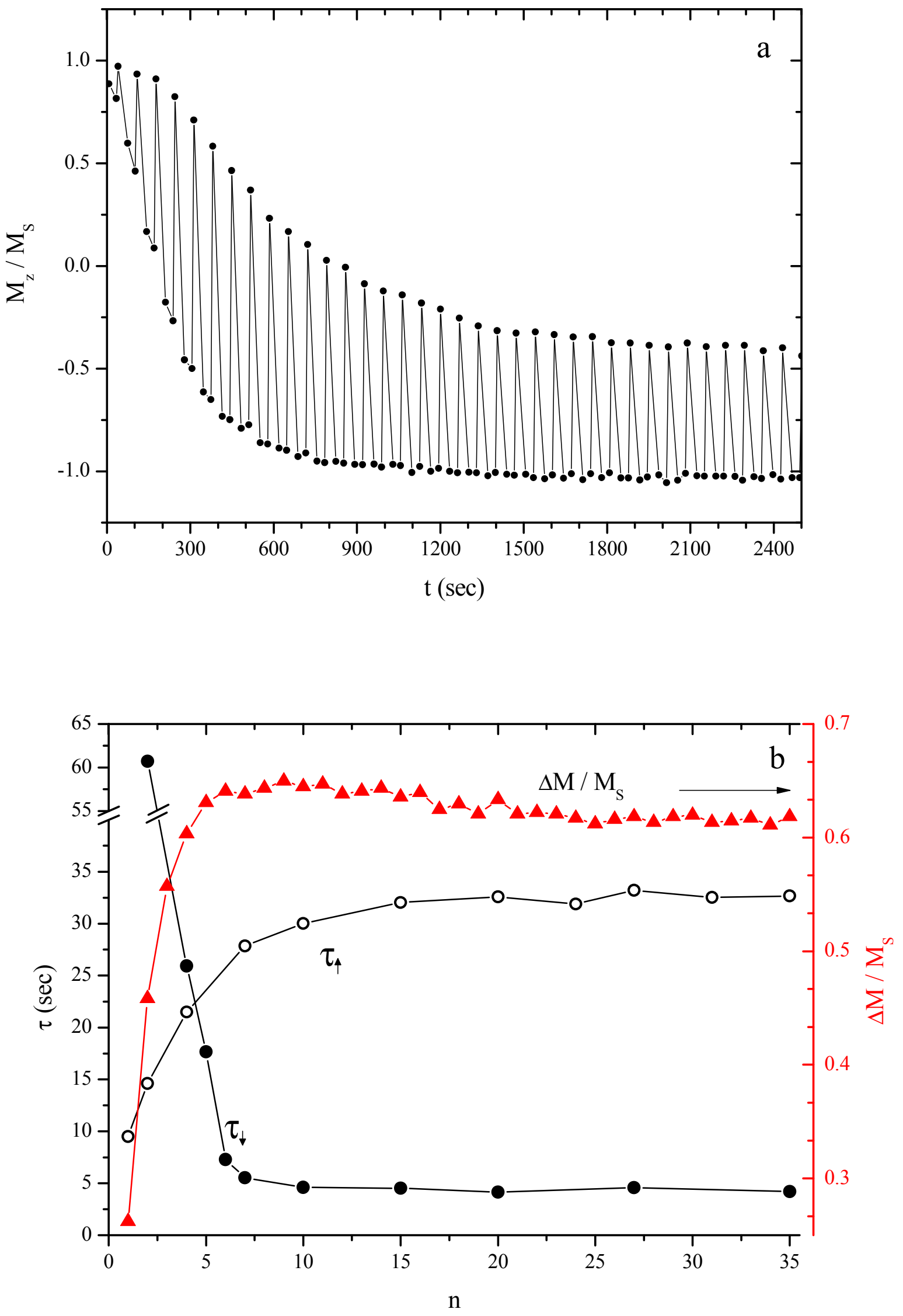

Figure 11

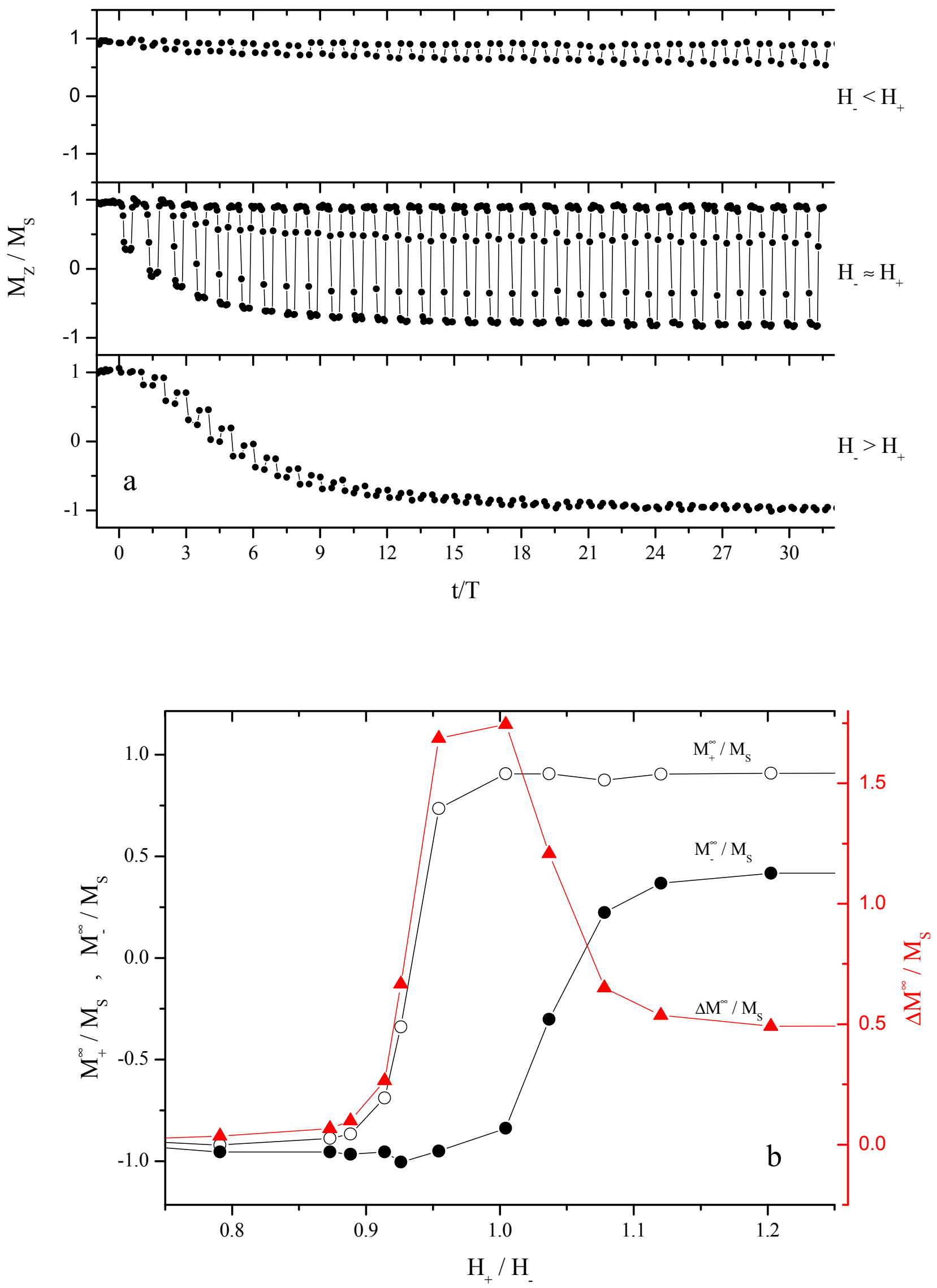

Figure 12

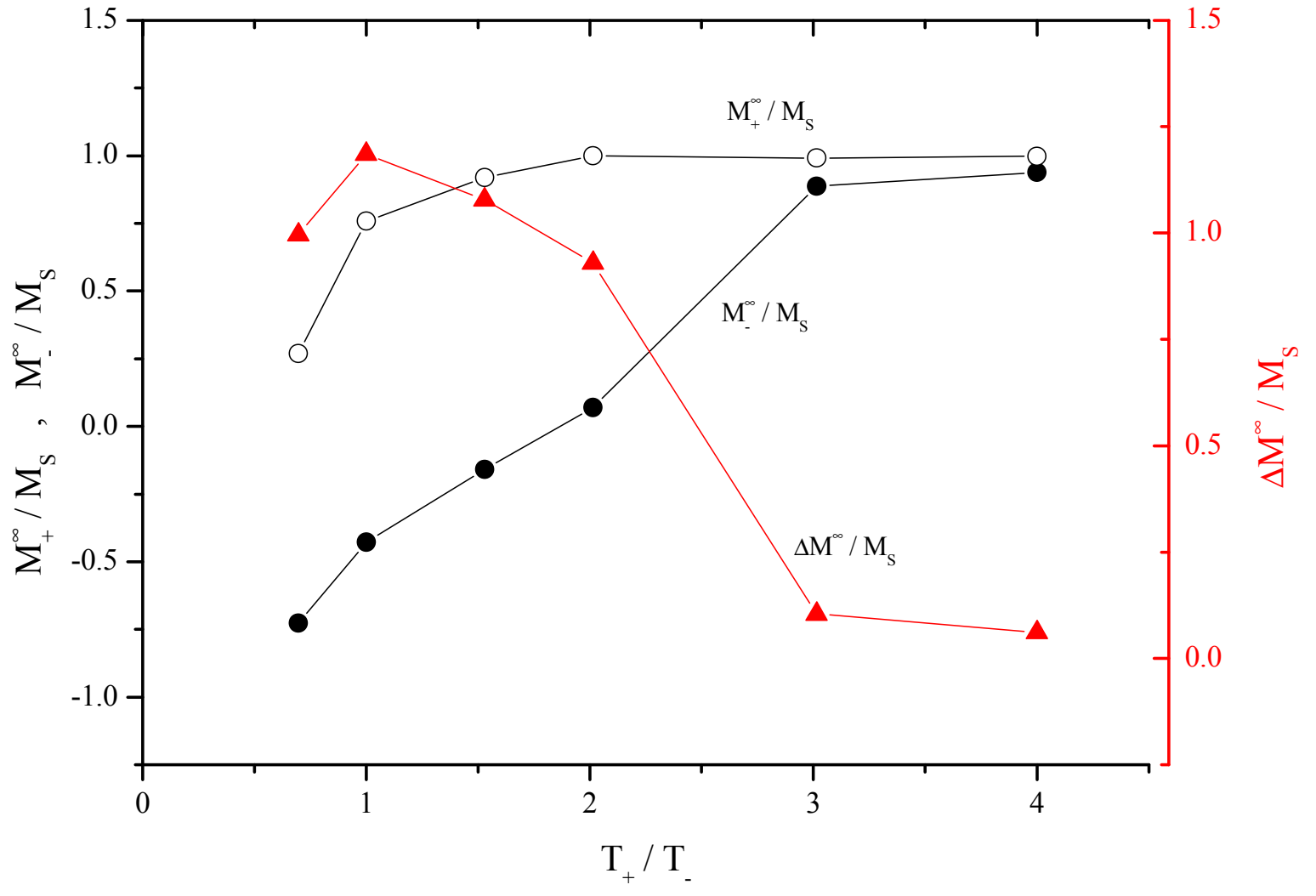